# Collaboration Analysis Using Deep Learning

Zhang Guo[1], Kevin Yu[2], Rebecca Pearlman[3], Nassir Navab[2,4], and
Roghayeh Barmaki[1,*]

[1] Department of Computer and Information Sciences, University of Delaware
[2] Chair for Computer Aided Medical Procedures and Augmented Reality, Technical University of Munich
[3] Department of Biology, Krieger School of Arts and Sciences, Johns Hopkins University
[4] Laboratory of Computational Sensing and Robotics, Johns Hopkins University

Correspondence to: Dr. Roghayeh Barmaki, E-mail: rlb@udel.edu

**ABSTRACT**

The analyses of collaborative learning process is one of the growing field of education research, which has many different analytic solutions. In this paper, we provided a new solution to improve the automated collaborative learning analyses using deep neural networks. Instead of using self-reported questionnaires, which are subject to bias and noise, we automatically extract group-working information by object recognition results using Mask R-CNN method. This process is based on detecting the people and other objects from pictures and video clips of the collaborative learning process, then evaluate the mobile learning performance using the collaborative indicators. We tested our approach to automatically evaluate the group-work collaboration in a controlled study of 33 dyads while performing an anatomy body painting intervention. The results indicate that our approach recognizes the differences of collaborations among teams of treatment and control groups in the case study; $F_{(1,32)} = 4.83$, $p < 0.05$. This work introduces new methods for automated quality prediction of collaborations among human-human interactions using computer vision techniques.

## 1. INTRODUCTION

The past few years have shown that collaborative learning is an effective educational strategy for the educators, who wanted to help those isolate learners join the group interaction, and who realized language skills they taught cannot make meaning without a good language environment and face-to-face practice. Many researches have acknowledged collaborative learning can improve students' learning motivation and increase knowledge retention at both theoretical and empirical level. With the big changes in student populations and the growth in the number of non-traditional learners, small group study format enables students to involve in the discussion, ask questions, and apply their knowledge to solve practical problems instead of individual learning or lecture-based learning.

However, it is undeniable that these benefits only work on the well-performed teams which have efficient collaborative activities during the learning process. Here comes the problem of collaboration learning: how can we find those "at-risk" groups who are likely to fail of group collaboration and need excessive help, and what is the sweet-spot for a well-functioning collaborative team.

Instructors in most universities set group working as part of their course assignments and assign even up to 100% of the final grade for group work. For example, in the affiliated university, several self-reported, and observational measures are used by course instructors to evaluate level of collaboration among teams of students using observation checklists, peer evaluation surveys, focus groups, and the grade of their group products. Though these methods are effective, but there are some constraints with these approaches, such as no objective measures to automatically evaluate the process of collaboration while students performing a group work. Recently, some scholars either established effective collaborative learning models or built reasonable standards for judging collaborative learning process based on self-reported survey data or collaboration system data. The core of collaborative learning analyses is building indicators based on the participants' performance during the entire learning process.

To encourage students in participation and help them better understand professional concepts and terminologies, different educational technologies and strategies are integrate into collaborative learning. By now, researchers have applied some well-functioned hand-held devices in education and such an education method is called mobile learning. Learners are able to use them to take lecture notes, take online courses, read e-books and have online group discussions. For human anatomy educators, they proposed body painting as an educational strategy, in which those complicate terminology is more visualized and easier to understand than the 2-D images on the textbooks.

What will happen if we combine these technologies and strategies together and apply them in collaborative learning process simultaneously? To answer this question, collabora-tive learning analyses are required to be performed on the collected data from both experimental groups and control groups.

What's more, consider using collaborative learning in a large laboratory course, where students are divided into hundreds of small groups. All groups are asked to learn the same objectives, yet they are assigned various learning methods and time spans to complete the learning task. Evaluation of participants' performance and the impact of educational technologies used in collaborative learning require an efficient collaboration data of the entire learning process. Using an automate data collection and analyses method instead of traditional survey will bring significant convenience. Many solutions have been proposed for collaborative learning analyses, such as Regression, Support Vector Machine, and Decision Tree [4, 20]. However, the data for those models were collected from class attendance, quizzes scores, reports, and course views, which could not directly reflect participants' performance during the learning activities. The most efficient data should be participants' actions and emotions during the process, which could be recorded via images or video clips. Also considering students, typically outnumber instructors in a large scale, instructors could enjoy more convenience collecting data using electronic devices than manually.

Traditional analyses methods cannot be directly applied on the image and video data. Yet object recognition methods can help us extract useful features from the raw data. Object recognition is a computer technology related to computer vision and image processing that deal with identifying instance of semantic objects of a certain class in digital images and videos [28]. Within object recognition domain, there are two main tasks: identification and categorization (also called classification). Image classification is to predict a set of labels to characterize the contents of an input image. Considering classification as a simple supervised problem, the learning module's input is an image and the output is a label of the class of one of the object in the image. The learning module is a binary classifier and it is trained with a couple of labeled examples. Object recognition builds on image classification but also allows image localize each object with a bounding box and even a segment. In practice, we need to consider about the size and composition of the training dataset for classification and the variability of each class for identification [25]. Those parameters and detection algorithms vary between different methods.

In this study, our goal of object recognition is to detect participants and their learning tools from the image data and then acquire features from participants' locations and finally recognize collaborative actions from the image data. Mask R-CNN (Region-based Convolutional Neural Network) is a Convolutional Neural Network (CNN) model. It has following outstanding features for object recognition: (1) it can not only detect the object but also generate a high-quality segmentation mask for each instance; (2) it is simple to train and flexible to use (3) it allows the system to run fast based on parallel heads (4) it can detect multiple objects in one image with high accuracy.

In this work, we evaluate the usefulness of the Mask-RCNN object recognition method in identifying collaborations in a tangible, body painting activity. We first introduce a new team-based anatomy education intervention using electronic hand-held devices, and then compare students' level of collaborative in experimental groups versus control groups using Mask R-CNN. Mask R-CNN trained with the COCO dataset was the method we used for object recognition. We will demonstrate the features acquired from the running results of Mask R-CNN could directly reflect students' collaboration performance.

## 2. RELATED WORK

Collaborative learning is a widely used education pattern featured by small group interaction and team-based evaluation metrics. Typically two or more participants are assigned in the same group and work for a common purpose which encourages them to learn via teamwork and cooperation [11]. With the computer-supported technology, many researchers even focus on online collaborative learning or long-distance collaborative learning based on the online system, multimedia, virtual reality [23], and mobile social applications [9]. For large number of students in one class, compared with the traditional individual learning or lecture-based learning in the classroom, effective collaborative interaction with peers promotes a positive attitude toward the subject matter for the students and increases student retention.

Using appropriate education technology is one of the main factors of facilitating learning and improving learning performance. To meet the 21st-century students' requirement, the term called "mobile learning" has been introduced by Jacob, and Issac, in 2008, which was based on the usage of mobile tablets for learning purposes [3]. According to the study of Cheung and Haw [6], mobile learning is an advanced strategy to motivate students engage into study and diversify the teaching model.

Learning the terminology and the concept like human anatomy challenges undergraduate students. Time is spent on somewhat inefficient learning by attending lectures, reading textbooks and reviewing 2-D images [2]. Besides using the mobile tablets, many successful efforts were proposed recently to improve active learning. Body painting [22], clay modeling [10], haptics simulation and models [19], and other models or methods used in education have added playful learning experiences.

The core of analyzing collaborative learning process is to build the cooperation indicators using the data gathered from group working and estimate the quality of the cooperation process [7]. Various approaches for analyzing group working interaction have been proposed. Soller and Lesgold [27] have provided an effective collaborative learning model as a framework, collected peer-to-peer conversation data using questionnaires and evaluated the active learning skills by computing the value for the following attributes: encourage, reinforce, rephrase, lead, suggest, elaborate, explain/clarify, justify, assert, information, elaboration, clarification, justification, and opinion. In 2007, they provided Hidden Markov Modeling approach to model collaborative learning [26]. Collazos et. al [7] divided the collaborative learning process into three stages and defined 5 of indica-tors to evaluate the collaborative learning process: applying strategies, intra-group cooperation, success criteria review, monitoring and performance. Later, in 2007 [8], they developed software tools to gather information automatically and used system-based measurement to improve their evaluation framework. With the widespread use of computers, computer supported collaborative work, has reshaped data collection and analytics in collaborative learning settings.

In this paper, we proposed to use object recognition techniques for collaborative learning analyses. Object recognition is a computer technology related to computer vision and image processing that deals with detecting instances of semantic objects of a certain class in digital images and videos [5]. The modern history of object recognition goes along with the development of deep learning techniques. Among all deep learning architectures, deep convolutional neural networks [15], widely used as image feature extractor, have the most impact on computer vision tasks. When dealing with the task, several popular deep neural networks stand out due to their high performance. AlexNet [15], based on LeNet [17], opened the new era due to the huge lead in ILSVRC, (2012). The R-CNN approach [13], a natural combination of heuristic region proposal method and ConvNet feature extractor, is able to detect the object and localize the object by bounding boxes. However, the problem with the R-CNN method is incredibly slow computation speed. Two improvements are introduced afterwards, Fast R-CNN [12] and Faster R-CNN [16]. Faster R-CNN uses the same algorithm as R-CNN to extract region proposals. The difference comes from RoIPool (RoI represents Region of Interesting) module which works by extracting a fixed-size window from the feature map and using these features to obtain the final class label and bounding box. It is effective and end-to-end trainable and the high speed makes it able to do real-time object recognition [24]. Mask R-CNN [14] extends Faster R-CNN for instance segmentation by adding a branch for predicting class-specific object mask, in parallel with the existing bounding box regressor and object classifier. The additional mask output comes from the finer spatial layout of the object. Mask R-CNN can be generalized to other tasks as well, such as human poses estimation, without changing the same framework. In this experiment, we use a pre-trained Mask R-CNN with the COCO [18] dataset as our object recognition approach.

## 3. METHOD

In this section, we explain the details of the case study, including the course in which the data were collected, and the specifications of the education technology used in the muscle painting activity and collaborative learning analyses using deep learning method.

### 3.1 Case Study

We conducted a between-subjects study while performing an anatomy learning intervention in a large laboratory course of General Biology offered by the Department of Biology at the Johns Hopkins University in spring 2018, which was enrolled by over 300 undergraduate students. Students were working in pre-assigned teams for the entire course to complete several anatomy lab exercises, including the muscle painting.

### 3.1.1 Muscle Painting Activity

During the muscle painting activity, pairs of students collaborate to paint 12 muscles of their body using painting supplies. The first student plays the role of a model, while her teammate, as a painter, locates the major upper-limb muscles using the human anatomy diagram in the lab manual [21], and then paints her upper limb with painting supplies. Afterwards, students switch roles, and the upper limb painter becomes model for lower limb. The goal of this activity is that students both get the knowledge of anatomy in a collaborative effort. The painting activity was upgraded for experimental group by using mobile tablet devices instead of the textbook for visualizing the musculoskeletal system. Figure 1 shows two settings of the study to complete the muscle painting activity.

**Figure 1: Case study setup for participants in pairs to complete the painting activity using either (a) textbook, or (b) tablet.**

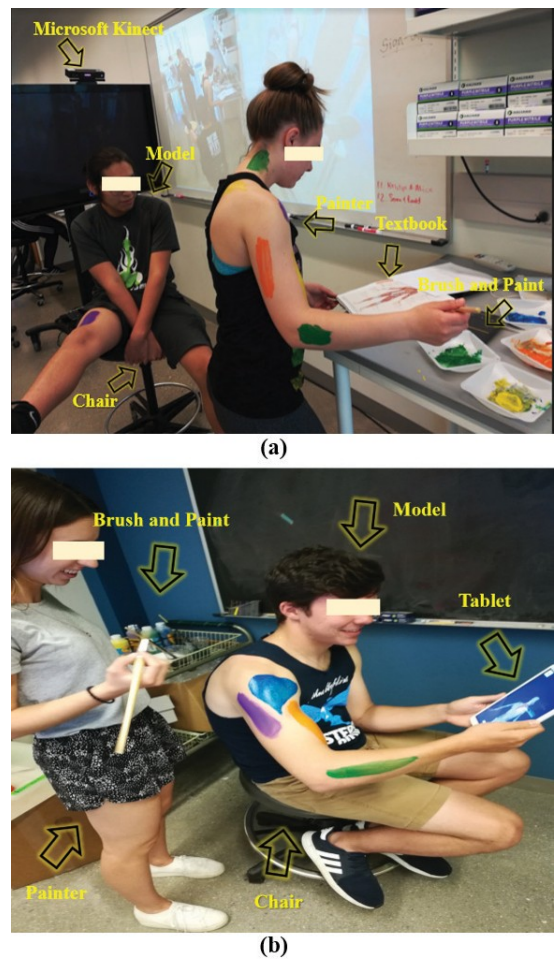

### 3.1.2 Procedure

An online flyer was sent to all students inviting them to participate in the study. The study was approved by the Institutional Review Board, and oral informed consent was obtained from each participant before the case study commenced. After consent, students completed an online pre-questionnaire (and a pre-test based on their assigned group) individually, and then entered the painting activity room with their preassigned teammates. Each team either used (1) textbook (control conditions), or (2) Tablet (experimental conditions) to complete the painting task. All students completed pre and post questionnaires before and after the intervention.

There was a mobile workstation for each laboratory room to capture videos and photos so that students' performance during the painting activities. The laboratory room was small and clean and participants were asked to stand close to the workstation in order to have appropriate image data.

## 3.2 Datasets

**Image data.** Image data is the pictures and video clips captured from the muscle painting scene every ten-seconds (Figure 2(a)). It showed the entire collaborative learning activities of each group from start to finish. Since we were going to use object recognition techniques to detect the people and their behaviors afterwards, only the group members and the tools they used for painting activity were captured in the image data. Based on different time spent on the painting activities, data for each group consist of 25 to 200 images. Each image file was name after the shooting time. When choosing the image data for object recognition, we focus on how close the team members were during the painting activity and treat such distance as a feature. In this study, we worked on groups of size two.

**COCO dataset.** COCO dataset is an online open source dataset and contains photos of 90 easily recognized common objects categories, including person, chair, desk, bottle, cell phone, book, etc. Over 2.5 million instances are labeled using per-instance segmentation in 328k images to aid in precise object localization [18]. COCO dataset can annotate instance-level segmentation mask, and can be used in both image classification and object recognition task for both iconic images and non-iconic images [18].

## 3.3 Object Recognition

**Mask R-CNN approach.** Mask R-CNN approach uses the same Region Proposal Network (RPN) stage as Faster R-CNN to generate bounding box for each candidate object and replaces RoIPool by a more accurate module RoIAlign to extract features from each box and show the bounding-box regression and classification [14] (Figure 3). By feeding the input image, a CNN feature extractor is able to extract image features which are called feature maps. Then then a CNN RPN will create RoIs which are the candidate object regions generated by RPN and ranked based on their score (how likely is the candidate object region could contain an object). Then the N (Faster R-CNN: $N$ = 2000; Mask R-CNN: $N$ = 300) regions with the highest scores are kept. Each of them will be warped into fixed dimension by RoIAlign and feed into three parallel branches: 2 Fully Connected (FC) layers as Faster R-CNN make classification and boundary box prediction and 2 additional convolution layers to build the mask. The top 100 detection boxes are kept and form a $100 \times L \times 15 \times 15$ tensor, where $L$ represents the number of classes in the training dataset (COCO dataset: $N$ = 90), and $15 \times 15$ represents the size of each predicted mask. The resized masks and the bounding boxes can be overlaid on the original input image as a transparent layer.

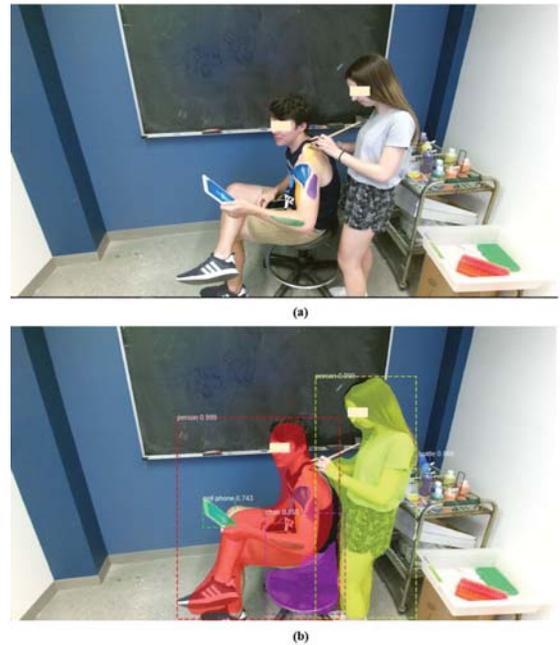

Figure 2: Collaborative learning using tablet in painting activity (a) original image data, and (b) object recognition using Mask R-CNN.

Using a pre-trained neural network by COCO dataset [1], we are able to apply Mask R-CNN to recognize the participants and the painting tools from the collaborative learning activities image data automatically as Figure 2. In our implementation, we focused on the following features of the object recognition results (Figure 3(b)): image file name, category name, bounding box coordinates and the score for each object. These features are the key points for the collaborative learning analyses and are stored in an output file in CSV format.

## 3.4 Collaborative Learning Analyses

To evaluate students' performance during the collaborative learning process, we build two indicators: Time on Task, and Level of collaboration in terms of proximity among team members.

**Time on Task.** We also calculated the time that students dedicated to perform the collaborative painting task. Previous work shows that increased time on task is a strong predictor of knowledge retention. We consider time on task as one of the indicators of student engagement as well. Previous activities indicated that un-engaged students just finish the activity in minimum possible time; so we articulate that students in treatment group may spend more time on task, and henceforth may engage more in the collaborative activity.

**Level of Collaboration.** In this muscle painting activity, collaboration was crucial in terms of how closely the participants work together. Especially close physical distance or proximity, and the amount of time students work together to perform the task was part of the learning process.

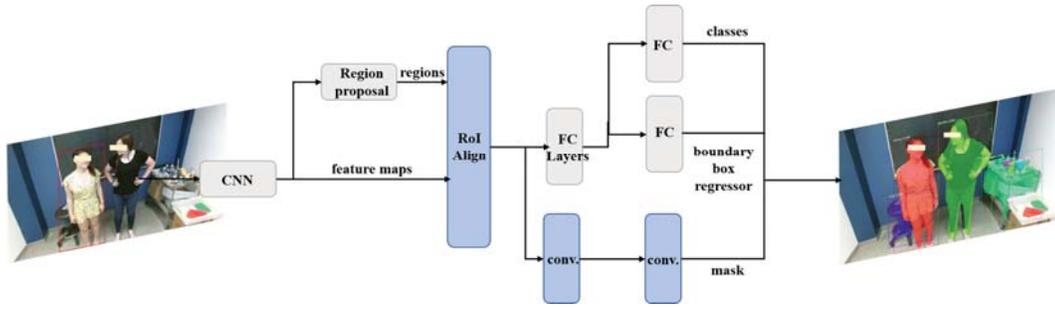

Figure 3: The Mask R-CNN framework for object recognition.

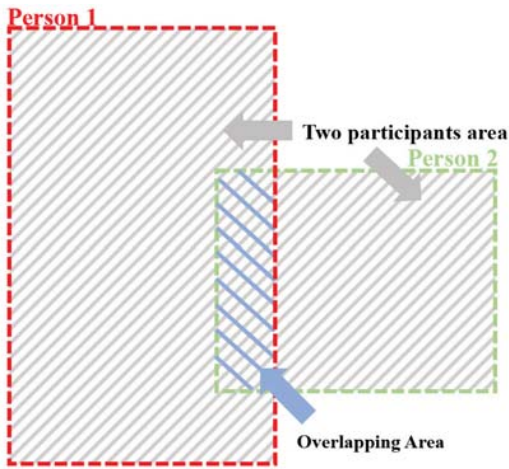

Figure 4: The overlapping ratio is equal to the overlapping area divided by the smaller participant area.

We measured this factor using Mask R-CNN approach, and named it level of collaboration. We hypothesize that teams of students who use the tablet, collaborate more with each other, and are within each ether's proximate distance more often in comparison to control group who use textbook. Once again, we want to highlight that our assumption may only fit for tangible work-group activities like body painting with pairs of students, and may not be generalized to other types of group-work activities which need distributed task allocations. Next we describe how we formulated the proximity of students using bounding boxes of study participants.

**Overlapping area and ratio.** The idea for computing overlapping area and ratio came from how close two participants were during the painting activity. Since there are lots of body contacts between painters and models while collaboration study, their bounding boxes may overlap. Using the coordinates of the bounding boxes of the person, we calculate the overlapping area (if their bounding boxes do not overlap at all, the overlapping area and ratio will be 0) and the whole area for two participants (Figure 4). The overlapping ratio is equal to the overlapping area divided by the smaller participant area. The reason of comparing the overlapping area with the smaller participant area is to eliminate the bias that one participant may have larger box than the other.

Get the percentage of the overlapping rate. For each group, compute the mean of all the overlapping rates in the object recognition results.

Time on Task. Time on task means the time slots (in second) each group spent in completing the painting activity, which is computed using the file names of the image data.

## 4. RESULTS

In this section, we report the findings from our case study. A summary of results is reported in Table 1. **Participants.**

Total of 66 (39 Female) students in 33 teams participated in this study. 17 out of 33 pairs were in treatment group (14 Female), and 16 dyads were in control group (25 Female).

**Level of Collaboration.** Among 33 teams, we calculated level of collaboration using the aforementioned method, and compared treatment versus control group. The results showed that there was a statistically significant difference between these two groups based on level of collaboration; $F_{(1,32)} = 4.83$, $p < 0.05$, Cohen's d = 0.77 (relatively large effect size).

**Time on Task.** Similarly, treatment group had a meaningful difference with control group based on time on task $F_{(1,32)} = 4.22$, $p < 0.05$, Cohen's d = 0.73 (relatively large effect size).

Table 1: Summary of the descriptive analytics for two groups of the study.

| Group | Level of Collaboration (%) [a] Mean (±SD) | Time on Task (seconds) Mean (±SD) |
|---|---|---|
| Treatment | 8.5 ± 2.16 | 629 ± 353 |
| Control | 6.3 ± 1.85 | 428 ± 170 |

[a] Level of collaboration was measured based on the percentage of overlapping area of dyads while working on the task.

## 5. DISCUSSION AND CONCLUSION

In this paper, we introduced a tangible, collaborative muscle painting activity. We then tested two collaborative indicators related to team's proximity and time on task. We introduced an approach for automating the process of recognizing level of collaboration among pairs of students us-ing deep neural networks. We tested our approach on the collected data from 33 teams (and 66 participants). The analysis showed that our approach is capable of recognizing differences in level of collaboration among students in treatment versus control groups. Both time on task and level of collaboration could successfully distinguish the differences among two groups of the study.

There are some approaches to improve this work. Based on using COCO dataset as training dataset, Mask R-CNN method can only detect the participants as person class objects instead of identify different participants in each group. So, all analyses is team-based, and for individual performance evaluation as one of our extended aims, we may not be able to use current standard.

In the future, we might change the training dataset and the training weight to achieve higher accuracy. Currently, we use COCO trainig dataset and weigths in this approach. We also aim to collect some additional multimodal features, such as emotion recognition, head and body poses to better understand level of collaborations among groups.